\documentclass{article}
\usepackage{bm}
\usepackage{amsmath}
\usepackage{eufrak}
\usepackage[usenames]{color}
\usepackage[utf8]{inputenc}
\usepackage{url}

\DeclareMathAlphabet{\mathpzc}{OT1}{pzc}{m}{it}
\newcommand{\ff}{\mathpzc{f}}
\newcommand{\dd}{\mathrm{d}}
\newcommand{\ii}{\mathrm{i}}
\newcommand{\eq}[1]{Eq.~\ref{eq:#1}}

\newcommand{\sect}[1]{Section \ref{sec:#1}}

\newcommand{\mAA}{{\mathrm \AA}}

\begin{document}                  % DO NOT DELETE THIS LINE

\title{Relationship between the atomic pair distribution function and small angle
scattering: implications for modeling of nanoparticles}

\author{
Christopher L. Farrow\\
\small Department of Physics and Astronomy, Michigan State University,\\ 
\small East Lansing, MI, 48824, USA
\and
Simon~J.~L. Billinge\\
\small Department of Applied Physics and Applied Mathematics, Columbia University,\\ 
\small New York, NY, 10027, USA \\
\small \texttt sb2896@columbia.edu}

\date{}
\maketitle                        % DO NOT DELETE THIS LINE

\begin{abstract}

Here we show explicitly the relationship  between the functions used in the
atomic pair distribution function (PDF) method and those commonly used in small
angle scattering (SAS) analyses.  The origin of the sloping baseline, $-4\pi
r\rho_0$, in PDFs of bulk materials is identified as originating from the SAS
intensity that is neglected in PDF measurements. The non-linear baseline in
nanoparticles has the same origin, and contains information about the shape and
size of the nanoparticles.

\end{abstract}

\section{Introduction}

The atomic pair distribution function (PDF) analysis of x-ray and neutron powder
diffraction is growing in popularity with the advent of nanoscience and
nanotechnology.  The technique is more than 70 years
old~\cite{debye;pz30,warre;b;xd90,egami;b;utbp03} and was originally applied almost
exclusively to the study of glass and amorphous
structures~\cite{warre;jpc34,warre;jacers36,frank;ac50,frank;prsla51,wrigh;gpc98}.
However, the approach is proving powerful in solving structure on the
nanoscale~\cite{billi;cc04}, where traditional crystallographic methods break
down~\cite{billi;s07}.  In particular, the study of the structure of discrete
nanoparticles using the PDF method has recently become a focus
\cite{mcken;n92,zhang;n03,gates;jpcb04,gilbe;s04,page;chpl04,korsu;jac05,petko;jmc05,bedfo;jpcc07,ehm;pd07,masad;prb07,pradh;cm07}.
The convergence of this new need with the availability of powerful sources of
high energy synchrotron x-rays and spallation neutrons and fast computing is
greatly expanding the power and applicability of the method.

The PDF, $G(r)$, is defined both as a function of the real-space pair density,
$\rho(r)$, and the reciprocal space scattering, $F(Q)=Q[S(Q)-1]$, as follows:
\begin{equation}
G(r) = \frac{2}{\pi}\int_0^\infty F(Q)\sin Qr \dd Q
\label{eq:grFromfTraditional}
\end{equation}
and
\begin{equation}
G(r)      = 4\pi r(\rho(r) - \rho_0),
\label{eq:grFromRhoBulk}
\end{equation}
where $\rho_0$ is the number density of the material
\cite{%
kaplo;pr65,% Kaplow
kaplo;pr68,% Kaplow
klug;b;xdpfpaam74,% Klug book
johns;jncs82,% Glass paper, includes A. C. Wright
soper;prl82,% Soper on water
korsu;jsc85,% Early RDF work by Korsunskiy
wrigh;jncs85,% A. C. Wright, glasses with neutrons
nanao;prb87,% Early crystal PDF with Egami
warre;b;xd90,% Warren, of course
egami;ferro91,% Egami, early methods paper
billi;prl94,% Early billinge @ Los Alamos
petko;prb98,% Petkov in Bulgaria
petko;jac98,% Petkov in Bulgaria
keen;jac01,% Keen's PDF lexicon
tucke;jac01a,% RMC for crystals
egami;b;utbp03,% Egami and Billinge
soper;jpcm07ii,% Soper on water
neder;pssc07,% Recent neder paper
billi;jssc08,% Latest billinge "big picture" paper
dinne;b;pdtap08,% Dinnebier and Billinge
gilbe;jac08% Latest gilbert theory paper
}.
Equation \ref{eq:grFromRhoBulk} in this form works well for bulk materials, but
the negatively sloping baseline, $-4\pi r\rho_0$, is no longer valid when the
PDF is calculated from finite-sized objects such as discrete nanoparticles
\cite{korsu;jac05}.  Motivated by the need for a rigorous definition of the form
of this baseline we rederive these equations here.  We show the correct form of
\eq{grFromRhoBulk} in a number of cases of practical interest such as discrete
nanoparticles and nanoparticle PDFs calculated from bulk models.  The important
distinction is provided by the small angle scattering (SAS) intensity and we
explicitly relate the commonly used PDF functions with commonly used results
from SAS.  We also show that the widely used pair of definitions for the PDF
above are actually incompatible with each other and that the definition
\eq{grFromfTraditional} does not give rise to \eq{grFromRhoBulk} but rather to
$R(r)/r$, where $R(r)$ is the radial distribution function.  This work therefore
resolves a long-standing ambiguity in the PDF literature.

Compared to the PDF of a bulk sample~\cite{levas;jcc07}, the PDF of a
nanoparticle is attenuated with increasing-$r$ by a function that is related to
the form of the nanoparticle \cite{guini;b;xdic63}. For simple shapes, such as
spheroids, spherical shells, rods and discs, this nanoparticle form factor can
be computed
analytically~\cite{rayle;prsl14,glatt;b;saxs82,thorp;unpub07,gilbe;jac08} and
integral equations exist for more complex shapes~\cite{kodam;aca06}. This lends
itself to a simple nanoparticle modeling procedure where the nanoparticle PDF is
calculated from the PDF of a bulk phase analogue by multiplying by the assumed
nanoparticle form factor~\cite{guini;b;xdic63,qiu;prl05,masad;prb07}.  The
approach is successful for extracting precise quantitative structural
information about the crystalline core of nanoparticles, including defects and
size-dependent bond-lengths~\cite{masad;prb07}, and this functionality has been
incorporated in the latest version of the PDF modeling software
PDFgui~\cite{farro;jpcm07}.  However, the method is not applicable when the
nanoparticle structure has no bulk phase analogue. This is the case in general,
for example in  nanoparticles with surface modifications~\cite{zhang;n03} and
inhomogeneous compositions such as core-shell nanoparticles~\cite{lizma;l96}. In
these cases, models of discrete nanoparticles must be applied.  As we discuss
below, this results in an ambiguity about the precise form of the measured
correlation function, and therefore how to calculate it.  Currently, this is
dealt with quite successfully in an \emph{ad hoc} way, as for example
in~\cite{korsu;jac05,neder;jpcm05,korsu;jac07,neder;pssc07}.  This paper
presents a rigorous definition of the form of the baseline in terms of the
nanoparticle form factor.

In Section \ref{sec:derivation} we rederive the equations giving rise to the PDF
to show the precise relationship between the measured correlation function in an
x-ray or neutron total scattering experiment and the underlying model.  In
Section \ref{sec:sas} we make explicit the link between the commonly used PDF
and small angle scattering equations.  This has implications for calculating
PDFs from discrete nanoparticle models for quantitative comparison with data,
which are discussed in \sect{methods}.  In \sect{sasft} we discuss the
conditions under which the PDF can be calculated in real space. Section
\ref{sec:summary} contains a brief summary.
% Chris' references. Saved this so we don't lose these.
% ~\cite{lizma;l96,rosen;ssr07} or those with surface or % geometrical
%restructuring~\cite{jadzi;s07}

\section{Derivation of the PDF equations}
\label{sec:derivation}

To understand the precise relationship between the commonly used PDF equations,
nanoparticle structures, and small angle scattering we rederive the PDF
equations from the beginning since subtle details of the derivation that are
often overlooked have a significant impact on discussion presented here.
Furthermore, the full derivation is not reproduced even in many textbooks on the
subject~\cite{warre;b;xd90,klug;b;xdpfpaam74,egami;b;utbp03,dinne;b;pdtap08} and
so these subtleties are not widely appreciated in the community.  We start from
the scattering amplitude from a set of $i$ atoms at points $\vec{r_i}$ in the
kinematical limit:
\begin{equation}
\begin{split}
\psi(\vec{Q}) &= \sum_i f_i(Q)e^{\ii\vec{Q}\cdot\vec{r_i}}\\
&= \sum_i \psi_i
\end{split}
\end{equation}

If the scattering from these atoms were totally incoherent the total intensity
would be the sum of the intensities from each atom,
\begin{equation}
\begin{split}
I_{inc}   &=\sum_{i} \psi_i^* \psi_i \\
          &=\sum_{i} f_i^*(Q)f_i(Q) \\
          &=\sum_{\alpha} N_\alpha f_\alpha^*(Q)f_\alpha(Q)\\
          &=N \sum_{\alpha} c_\alpha f_\alpha^*(Q)f_\alpha(Q)\\
          &=N \langle f^2 \rangle,
\label{eq:iinc}
\end{split}
\end{equation}
where the sum over $\alpha$ is now over the different species of atoms in the
sample with $N$ being the total number of atoms, $N_\alpha$ being
the number of atoms of type $\alpha$ and where the concentration of species
$\alpha$ is $c_\alpha = N_\alpha/N$.
Similarly, we can define the sample-averaged scattering power, $\langle f
\rangle = \sum_\alpha c_\alpha f_\alpha$ and
\begin{equation}
\begin{split}
\langle f\rangle ^2 & = \frac{1}{N^2} \sum_{ij} f_j^* f_i \\
    & = \sum_{\alpha\beta} c_\alpha c_\beta f_\alpha^* f_\beta.
\end{split}
\end{equation}

The full coherent scattering intensity is given by $\psi^*\psi$ which is
\begin{equation}
\label{eq:Icoh}
\begin{split}
I_c &= \sum_i\sum_j f_j^*f_i e^{\ii\vec{Q}\cdot (\vec{r_i}-\vec{r_j})}\\
    &= \sum_{i,j} f_j^*f_i e^{\ii\vec{Q}\cdot\vec{r_{ij}}}.
\end{split}
\end{equation}
Here we have dropped the $Q$-dependence of the atomic scattering factors to
simplify the notation, but the $f$s are understood to retain their
$Q$-dependence.   We can separate out the self-scattering, $i=j$, for which
$\vec{r_{ij}}=0$:
\begin{equation}
\begin{split}
I_c &= \sum_i f_i^*f_i + \sum_{i\neq j} f_j^*f_i e^{\ii\vec{Q}\cdot \vec{r_{ij}}}\\
    &= N\langle f^2\rangle + \sum_{i\neq j} f_j^*f_i e^{\ii
    \vec{Q}\cdot \vec{r_{ij}}},
\label{eq:icoh}
\end{split}
\end{equation}
where we have used \eq{iinc}, resulting in an expression for the discrete scattering intensity for $i\ne j$ as
\begin{equation}
\begin{split}
I_d &= I_c - N\langle f^2\rangle\\
    &= \sum_{i\neq j} f_j^*f_i e^{\ii \vec{Q}\cdot \vec{r_{ij}}}.
\end{split}
\end{equation}

We want an expression for the total scattering structure function, $S(\vec{Q})$,
which is defined as $\frac{I_c}{N\langle f\rangle^2} - \frac{\langle (f -
\langle f\rangle)^2\rangle}{\langle f\rangle^2}$.  The second term in this
definition is the Laue monotonic diffuse scattering that comes about because of
the imperfect cancellation of intensity at the destructive interference
condition when atomic sites are occupied by atoms of different scattering
strength. It results in a monotonic incoherent background even in the case of
perfectly coherent scattering. 

To get $S(\vec{Q})$ from \eq{icoh} we therefore must normalize by the total
number of scatterers, $N$,
\begin{equation}
\begin{split}
\frac{I_c}{N} &= \langle f^2\rangle + \frac{1}{N}\sum_{i\neq j} f_j^*f_i e^{\ii \vec{Q}\cdot \vec{r_{ij}}}.
\end{split}
\end{equation}
Subtracting the normalized self scattering term to get
\begin{equation}
\frac{I_c}{N} - \langle f^2\rangle =  \frac{1}{N}\sum_{i\neq j} f_j^*f_i e^{\ii
\vec{Q}\cdot \vec{r_{ij}}},
\end{equation}
and then normalizing by $\langle f\rangle^2$, we obtain
\begin{equation}
\frac{I_c}{N\langle f\rangle^2} - \frac{\langle f^2\rangle}{\langle f\rangle^2} =  \frac{1}{N\langle f\rangle^2}\sum_{i\neq j} f_j^*f_i e^{\ii \vec{Q}\cdot \vec{r_{ij}}}.
\end{equation}
Thus,
\begin{equation}
\label{eq:sofqdebye}
\begin{split}
S(\vec{Q})-1 &=\frac{I_c}{N\langle f\rangle^2} - \frac{\langle f^2\rangle}{\langle f\rangle^2}\\
&=\frac{I_d}{N\langle f\rangle^2}\\
&= \frac{1}{N\langle f\rangle^2}\sum_{i\neq j} f_j^*f_i e^{\ii\vec{Q}\cdot \vec{r_{ij}}}.
\end{split}
\end{equation}
This expression yields precisely $S(\vec{Q})-1$ in terms of scattering from
atoms in our sample.

For an isotropic sample, e.g., a powder of crystals or nanoparticles, we assume
there to be a crystallite with every orientation with equal probability and we
can take an orientational average.  Place the $\vec{Q}$ along $z$ so that we can
express $\vec{Q}\cdot \vec{r_{ij}}= Q r_{ij}\cos \theta$.  Then the
orientational averaging means that $\theta$ takes all values with equal
probability.  The sample-averaged intensity for a pair of atoms will therefore
be
\begin{equation}
\begin{split}
\label{eq:avgexp}
\overline{e^{\ii \vec{Q}\cdot \vec{r_{ij}}}}
&=\frac{\int_0^{2\pi} \dd\phi \int_0^{\pi}\dd\theta e^{\ii Qr_{ij}\cos\theta}r_{ij}^2\sin\theta}
{\int_0^{2\pi} \dd\phi \int_0^{\pi}\dd\theta r_{ij}^2\sin\theta}\\
&=\frac{-2\pi r_{ij}^2\left[ e^{\ii Qr_{ij}\cos\theta}\right]_0^{\pi}}{4\pi
r_{ij}^2 \ii Qr_{ij}}\\
&=\frac{ \left[ e^{\ii Qr_{ij}}-e^{-\ii Qr_{ij}}\right]}{2\ii Qr_{ij}}\\
&=\frac{ \sin (Qr_{ij})}{Qr_{ij}}.
\end{split}
\end{equation}
Using this in \eq{Icoh} gives the average coherent scattering intensity, as
expressed originally by Debye~\cite{debye;ap15}.  From this we get the total
scattering structure function for an isotropic sample,
\begin{equation}
\begin{split}
S(Q)-1 &= \frac{1}{N\langle f\rangle^2}\sum_{i\neq j} f_j^*f_i \frac{ \sin (Qr_{ij})}{Qr_{ij}}.
\label{eq:sqminus1}
\end{split}
\end{equation}
Thus, the reduced total scattering structure function, $F(Q)=Q[S(Q)-1]$, is
\begin{equation}
\label{eq:fqdebye}
\begin{split}
F(Q) &= \frac{1}{N\langle f\rangle^2}\sum_{i\neq j} f_j^*f_i \frac{ \sin (Qr_{ij})}{r_{ij}}.
\end{split}
\end{equation}

It is convenient at this point to get rid of the $Q$-dependence of the x-ray
form-factors.  They are assumed to be isotropic so depend only on $Q$ and not
$\vec{Q}$, which is a good approximation for scattering from core electrons
especially.  Write $f(Q)=f(0)\tilde{f}(Q)$, where $\tilde{f}(Q)$ has value 1 at
$Q=0$ and contains the $Q$-dependence of the form-factor and $f(0)\approx Z$,
where $Z$ is the atomic number that scales the form factor.  The
Morningstar-Warren approximation~\cite{warre;jacers36} is that the $Q$-dependent
part of the form factors can be well approximated by an average $Q$-dependence,
$\overline{\tilde{f}(Q)}=\frac{1}{N_{species}}\sum_\alpha c_\alpha
\tilde{f}_\alpha(Q)$.  In this case the $Q$-dependence,
$\overline{\tilde{f}(Q)}^2$, comes out of the double sums in \eq{fqdebye} on the
top and the bottom and cancels out.  The $f$s that remain are $Q$-independent,
and normally replaced by the atomic number (modified by any anomalous scattering
factors).  The same result holds for neutron scattering where the $f$s are
replaced by coherent neutron scattering lengths, $b$.  These have no
$Q$-dependence and therefore the approximate method for removing the
$Q$-dependence is not needed.

Now we want to consider the inverse Fourier transform of $F(Q)$. Because $F(Q)$
is an even function, we use the sine-Fourier transform,
\begin{equation}
\label{eq:fofr}
\begin{split}
\ff(r) &= \frac{2}{\pi}\int_{0}^\infty F(Q)\sin(Qr) dQ.
\end{split}
\end{equation}
We choose the $\frac{2}{\pi}$ prefactor so that the direct sine transform
has a prefactor of $1$.  This is precisely the definition of the PDF in
\eq{grFromfTraditional}. From this we get
\begin{equation}
\begin{split}
\label{eq:frdebye}
\ff(r) &= \frac{2}{\pi}\int_{0}^\infty \frac{1}{N\langle f\rangle^2} \sum_{i\neq j} f_j^*f_i \frac{\sin(Qr_{ij})}{r_{ij}}\sin(Qr) dQ\\
&= \frac{2}{\pi N\langle f\rangle^2}\sum_{i\neq j} \frac{f_j^*f_i}{r_{ij}} \int_{0}^\infty \sin (Qr_{ij})\sin(Qr) dQ\\
&= \frac{1}{N\langle f\rangle^2}\sum_{i\neq j} \frac{f_j^*f_i}{r_{ij}} \,[\delta (r-r_{ij})-\delta (r+r_{ij})]\\
&= \frac{1}{r N\langle f\rangle^2}\sum_{i\neq j} {f_j^*f_i} \,[\delta (r-r_{ij})-\delta (r+r_{ij})]
\end{split}
\end{equation}
which, if we confine ourselves to the positive axis only, is
\begin{equation}
\begin{split}
\ff(r)&= \frac{1}{r N\langle f\rangle^2}\sum_{i\neq j} {f_j^*f_i} \,\delta
(r-r_{ij}).
\end{split}
\end{equation}

We can interpret $\ff(r)$ in terms of the radial distribution function (RDF).
The RDF, denoted $R(r)$, is defined for an elemental system such that for an
arbitrary atom $i$ at the origin, $R_i(r)\dd r$ gives the number of atoms in a
shell of thickness $\dd r$ at a distance $r$ from that atom and the total RDF is
the average of the partial RDFs over each atom taken at the origin. Thus, the
integral of the RDF between two bounds gives the number of atomic pairs per atom
with separation within those bounds.  Equation \ref{eq:frdebye} yields this
behavior if we multiply by $r$.  For a solid with $\alpha$ atomic species we get
\begin{equation}
\begin{split}
\int_a^b R(r)\dd r &= \int_a^b\frac{1}{N\langle f\rangle^2}\sum_{i\neq j} {f_j^*f_i} \,\delta
(r-r_{ij})\dd r \\
&= \frac{1}{ N\langle f\rangle^2}\sum_{i}\sum_{j\in S} {f_j^*f_i}\\
&= \frac{1}{\langle f\rangle^2}\sum_\alpha c_\alpha f_\alpha \sum_{j \in
S} {f_j^*},
\end{split}
\end{equation}
where $S$ is the set of atoms with distance from atom $i$ greater than $a$ and
less than $b$.  In the case of just one atomic species, this reduces to
\begin{equation}
\begin{split}
\int_a^b R(r)\dd r &= \frac{f^2}{f^2}\sum_{j\in S}1\\
&= N_a,
\end{split}
\end{equation}
as required.  Thus,
\begin{equation}
\label{eq:frrho}
\begin{split}
\ff(r)&= \frac{R(r)}{r} \\
    &= 4\pi r\rho(r).
\end{split}
\end{equation}
The second expression comes from the relationship between the RDF and the pair
density.  The pair density is defined such that $\int \dd r \dd\phi \dd\theta
r^2 \sin(\theta) \rho(r) = \int R(r) \dd r$, so that $4\pi r^2\rho(r) = R(r)$.
Comparing Equations~\ref{eq:fofr} and~\ref{eq:frrho} we see that the definition
\eq{fofr} does \emph{not} yield $G(r)$ (\eq{grFromRhoBulk}) and strictly
$\frac{2}{\pi}\int_0^\infty F(Q)\sin Qr \dd Q = R(r)/r \ne G(r).$  However, we
see below that in practice it is $G(r)$ and not $R(r)$ that is obtained
experimentally in most cases.

Finally, reordering \eq{frrho} we find
\begin{equation}
\label{eq:rho}
\begin{split}
\rho (r)&= \frac{\ff (r)}{4\pi r } \\
    &= \frac{1}{4\pi r^2 N\langle f\rangle^2}\sum_{i\neq j} {f_j^*f_i} \,\delta
(r-r_{ij}).
\end{split}
\end{equation}

In reality, $I_{c}(Q)$ is measured down to a minimum $Q$ due to the experimental
setup. This means that in general the forward scattering contributions are lost.
We will consider the impact of this on the measured real-space function
$\ff(r)$.  We rewrite the expression for the experimental $\ff(r)$ as
\begin{equation}
\begin{split}
\label{eq:sasgofr}
\ff(r;Q_{min}) &= \frac{2}{\pi} \int_{Q_{min}}^{\infty} F(Q) \sin(Qr) \dd Q\\
     &= 4\pi r \rho(r) - \frac{2}{\pi} \int_0^{Q_{min}}F(Q) \sin(Qr) \dd Q.
\end{split}
\end{equation}
Of course, we have a finite $Q_{max}$ as well, but this will be disregarded
during the following discussion, as the effects are well
understood~\cite{toby;aca92}.

\section{Low angle scattering intensity}
\label{sec:sas}

We will now consider a number of explicit examples to understand how the missing
forward scattering affects the measured $\ff(r;Q_{min})$.

Consider a bulk material of uniform density and infinite extent.  Since it has
no internal structure there is perfect cancellation of all the discrete
scattering intensity, $I_d$, everywhere except at $Q=0$ since for all
wavelengths it is always possible to find a pair of volume elements where the
scattering is exactly out of phase with each other and cancels.  The definition
of $I_d$ in terms of a double sum over atoms is not entirely appropriate for
this case (since there are no atoms!), but it gives an intuitive  feeling about
the behavior.
%and we can define an effective $N$ in terms of the density and
%volume, $N=\rho_0 V$:
\begin{equation}
I_d(Q) = (N^2-N)\langle f \rangle^2 \delta(Q),
\label{eq:Iinfhom}
\end{equation}
which is a delta-function spike at $Q=0$ sitting on a zero background.  The
integrated intensity in the delta-function scales like the volume of the sample
squared.  Strictly speaking the sample has to be infinite in extent, and
therefore $N = \infty$, to get a perfect delta-function.  From
Eqs.~\ref{eq:sofqdebye} and~\ref{eq:Iinfhom} we see that $S(0)-1=N-1 \approx N$
since $\lim_{Q \rightarrow 0} \sin Qr/ Qr = 1$.

Now we do this more formally.  Assume the sample has a uniform number density,
$\rho_0$.  Since the scattering length is defined per atom, the scattering
amplitude of a volume element $\dd \vec{r}$ at position $\vec{r}$ is $\psi=
\rho_0\langle f\rangle e^{\ii \vec{Q}\cdot\vec{r}}\dd\vec{r}$.  For an infinite
crystal, the intensity is given by
\begin{equation}
I_c(\vec{Q}) = \rho_0^2\langle f\rangle^2 \int \int e^{\ii
\vec{Q}\cdot(\vec{r}-\vec{r^\prime})} \dd \vec{r} \dd \vec{r^\prime},
\end{equation}
where the integrals are over all space, which gives a delta-function at $Q=0$.

If the material is finite in extent then the integrals are finite.  To evaluate
the integrals, we can define a shape function $s(\vec{r})$ such that inside the
shape $s=1$ and outside the shape, $s=0$. For such a material,
\begin{equation}
I_c(\vec{Q}) = \rho_0^2\langle f\rangle^2 \int \int
s(\vec{r})s(\vec{r^\prime)}e^{\ii \vec{Q}\cdot(\vec{r}-\vec{r^\prime})} \dd
\vec{r} \dd \vec{r^\prime}.
\end{equation}
Let us do a change of variables so that $\vec{r^{\prime\prime}} =
\vec{r}-\vec{r^\prime}$, and $\dd \vec{r^{\prime\prime}} = \dd \vec{r}$, in
which case we have
\begin{equation}
\label{eq:Ishape}
\begin{split}
I_c(\vec{Q}) &= \rho_0^2\langle f\rangle^2 \int \int
s(\vec{r^\prime})s(\vec{r^\prime}+\vec{r^{\prime\prime}})e^{\ii \vec{Q}\cdot\vec{r^{\prime\prime}}} \dd \vec{r^\prime} \dd \vec{r^{\prime\prime}}\\
 &= \rho_0^2\langle f\rangle^2  \int \dd \vec{r^{\prime\prime}}
 e^{\ii \vec{Q}\cdot\vec{r^{\prime\prime}}}\int
 s(\vec{r^\prime})s(\vec{r^\prime}+\vec{r^{\prime\prime}})\dd \vec{r^\prime}.
\end{split}
\end{equation}
The second integral is a self convolution, or autocorrelation function, of the
shape function. Let us define
\begin{equation}
\label{eq:gammadef}
\gamma_0(\vec{r})=\frac{1}{V}\int s(\vec{r^\prime})s(\vec{r^\prime}+\vec{r})\dd
\vec{r^\prime},
\end{equation}
where $V = \int s(\vec{r}) \dd \vec{r}$ is the volume defined by the shape
function. This $\gamma_0(\vec{r})$ is the characteristic function of the
shape~\cite{guini;b;sas55}, and has been called the nanoparticle form factor in
the PDF literature~\cite{qiu;prl05,kodam;aca06,masad;prb07}.  Defined as such,
$\gamma_0(\vec{r})$ has the following properties:
\begin{equation}
\label{eq:gammaprops}
\begin{split}
\gamma_0(0) &=1 \\
\int \gamma_0(\vec{r}) \dd \vec{r} &= V
\end{split}
\end{equation}
This definition, and a convenient dropping of the double-primes gives
\begin{equation}
\begin{split}
I_c(\vec{Q}) &= \rho_0^2\langle f\rangle^2  V \int
\gamma_0(\vec{r})e^{\ii \vec{Q}\cdot\vec{r}}\dd \vec{r}.
\end{split}
\end{equation}

In analogy with the discrete case we want to convert this to $S(\vec{Q})-1 =
\frac{I_c}{N\langle f\rangle^2} - \frac{\langle f^2 \rangle}{\langle
f\rangle^2}$, and using the fact that $N=\rho_0 V$ we get
\begin{equation}
\label{eq:sqgamma}
\begin{split}
S(\vec{Q})-1
    &= \frac{1}{N\langle f\rangle^2}\rho_0^2\langle f\rangle^2  V \int \gamma_0(\vec{r})e^{\ii \vec{Q}\cdot\vec{r}}\dd \vec{r} - \frac{\langle f^2 \rangle}{\langle f\rangle^2}\\
    &= \rho_0 \int \gamma_0(\vec{r})e^{\ii \vec{Q}\cdot\vec{r}}\dd \vec{r}- \frac{\langle f^2 \rangle}{\langle f\rangle^2}.
\end{split}
\end{equation}
The second term, $\frac{\langle f^2 \rangle}{\langle  f\rangle^2}$, is very
small compared to the first term. It is order unity, where the first term scales
as $N=\rho_0 V$, and can safely be ignored in most cases.

Now we want to take the orientational average.  This must be done with care as,
in general, the orientation of the nanoparticle shape and the underlying
structure are correlated.  For example, the morphology of the particles (plates
or needles) depends on easy growth directions of the underlying structure.  As
discussed by Gilbert~\cite{gilbe;jac08}, this means that the shape function and
the internal structure of the particle are not, in general, separable and it is
not correct to get the scattered intensity by convolving the reciprocal-space
intensity with the Fourier transform of the characteristic function.  Things are
greatly simplified in the case where the underlying structure, or the
nanoparticle shape, or both, are isotropic, or approximately so.  Then we can
denote the angle-averaged characteristic function as
\begin{equation}
\label{eq:gammaavg}
\begin{split}
\overline{\gamma_0(\vec{r})} = \gamma_0(r) = \frac{
            \int \dd \phi \int \dd \theta \sin(\theta) r^2 \gamma_0(\vec{r})
            }
            {
            \int \dd \phi \int \dd \theta r^2 \sin(\theta)
            }.
\end{split}
\end{equation}
Using this and \eq{avgexp} we get
\begin{equation}
\label{eq:sqavg}
\begin{split}
S(Q)-1 &= \rho_0 \int_{0}^\infty \dd r \int_0^{2\pi} \dd\phi
            \int_0^{\pi}\dd\theta \overline {\gamma_0(\vec{r})
            e^{\ii \vec{Q}\cdot\vec{r}}} r^2\sin\theta \\
       &= \rho_0 \int_{0}^\infty \dd r \int_0^{2\pi} \dd\phi
            \int_0^{\pi}\dd\theta \overline {\gamma_0(\vec{r})}
            \overline{e^{\ii \vec{Q}\cdot\vec{r}}} r^2\sin\theta \\
    &= \rho_0 \int_{0}^\infty \dd r \int_0^{2\pi} \dd\phi \int_0^{\pi}\dd\theta \gamma_0(r)\frac{\sin Qr}{Qr}r^2\sin\theta \\
    &= \rho_0 \int_{0}^\infty \gamma_0(r)\frac{\sin Qr}{Qr} 4 \pi r^2 \dd r.
    \end{split}
\end{equation}
Since the particles have no preferred orientation in space, we have broken the
average of the product in the first line into the product of the averages.  This
gives
\begin{equation}
\label{eq:fqgamma}
\begin{split}
F(Q) &=   \int_{0}^\infty 4\pi \rho_0 r \gamma_0(r)\sin(Qr)\dd r.
\end{split}
\end{equation}
Noting that this is the direct sine-Fourier transform, we take the inverse
transform to get
\begin{equation}
\label{eq:fhomsolid}
\begin{split}
\ff_u(r) &=  \frac{2}{\pi} \int_{0}^\infty F(Q) \sin (Q r) \dd Q\\
&= 4\pi\rho_0 r \gamma_0(r),
\end{split}
\end{equation}
where the subscript $u$ indicates that this result is for a solid of uniform
density distribution, $\rho_0$.

Next we consider a macroscopic crystal. The difference in $F(Q)$ is at higher
$Q$ where, instead of complete cancellation of all the discrete intensity it
appears at distinct reciprocal lattice points as sharp Bragg peaks. Importantly,
in the region of $Q$ below the first Bragg peak, the distinct scattering is zero
except at very low-$Q$ where small angle scattering region is reached.  The
\emph{small angle} scattering intensity, $I_{sas}$ from the crystal is identical
to that from the solid with uniform density: $I^{sas}_u = I^{sas}_{crystal}$.
The small and wide angle scattering regions are well separated in $Q$ and
$I_{sas}$ decays to zero before $Q_{min}$ is reached in the crystal.  Thus,
\begin{equation}
\label{eq:fsas}
\begin{split}
\ff_{sas}(r) &= \frac{2}{\pi} \int_0^{Q_{min}}F(Q) \sin(Qr) \dd Q  \\
&= \ff_u(r)
\end{split}
\end{equation}
and therefore
\begin{equation}
\begin{split}
\frac{2}{\pi} \int_0^{Q_{min}}F(Q) \sin(Qr) \dd Q = 4\pi \rho_0 r\gamma_0(r).\\
\label{eq:fsasgamma}
\end{split}
\end{equation}

We are now in a position to understand in detail the nature of the measured PDF
$\ff(r;Q_{min})$.  Substituting \eq{fsasgamma} into \eq{sasgofr} we get
\begin{equation}
\label{eq:gofrgamma}
\ff(r;Q_{min})=4\pi r \rho (r) - 4\pi r \rho_0 \gamma_0(r).
\end{equation}
This is similar to the definition of the PDF from \eq{grFromRhoBulk}, except
that $\gamma_0(r)$ appears in the sloping baseline term.

\section{Calculating $\ff(r;Q_{min})$ from models}
\label{sec:methods}

We can now consider the calculation of measured PDFs using \eq{gofrgamma} in a
number of interesting limits.

\subsection{Calculating in real-space for bulk crystals}
\label{sec:calcBulk}

In the case of bulk crystals, the region of interest in the PDF is
usually $r \ll D$, $D$ being the smallest dimension of the crystal. In this
region, $\gamma_0(r) \approx 1$.  Thus,
\begin{equation}
\begin{split}
\label{eq:frbulk}
\ff(r;Q_{min})&=G(r)\\
&= 4\pi r (\rho_{bulk} (r) - \rho_0),
\end{split}
\end{equation}
which is the familiar definition of $G(r)$ in \eq{grFromRhoBulk}.  The pair
density function, $\rho_{bulk} (r)$, is calculated from a model with periodic
boundary conditions \cite{billi;b;lsfd98,proff;jac99}, or from a box of atoms
that is much larger in extent than the range of $r$ of interest
\cite{mcgre;ms88}, using \eq{rho}.  The
average number density $\rho_0$ is given by the number of atoms per unit volume,
which in the case of crystals is the number of atoms in the unit cell divided by
the unit cell volume.

Two approaches are typically taken to account for thermal and zero-point motion
of atoms.  One approach is to assume that these motions are well approximated by
a Gaussian probability distribution, in which case the delta-functions may be
convoluted by Gaussians of finite width to represent the motion.   If the motion
is anisotropic, Gaussian distributions with different widths in different
directions may be used.  Because the function being  convoluted is a
delta-function, from a practical perspective \eq{frbulk} is simply modified so
that a Guassian of the appropriate width is added instead of a
delta-function~\cite{proff;jac99,thorp;b;fstpbas02}.  In models with many
thousands of atoms, which is sometimes the case for reverse Monte Carlo methods,
the probability distributions can be built up from the static ensemble itself
and no convolution is carried out.  In this case no presumption of Gaussian
dynamics is made.  In practice the Guassian approximation works very well in
most cases, and deviations from Gaussian behavior can be accounted for by
introducing disorder in the models, which is a convenient way of separating the
harmonic and non-harmonic contributions to the structure.  The effects of
correlated dynamics are accounted for using an $r$-dependent Gaussian
broadening~\cite{jeong;jpc99,thorp;b;fstpbas02,jeong;prb03}.

A convolution is also often carried out to account for the termination effects
of the Fourier transform.  For example, if the data are simply terminated at
$Q_{max}$, as is often the case when data are available to high-$Q$, the model
PDF must be convoluted with a sinc function~\cite{egami;b;utbp03}.  From
Eqs.~\ref{eq:fofr} and~\ref{eq:frbulk}, it is clear that the function that must
be so convoluted is  $\ff(r;Q_{min})$.  This may be done with an integral
directly in real-space, but it is often quicker to use a fast Fourier transform
and do the convolution as a product in reciprocal space taking advantage of the
convolution theorem.  Note that, because of the sloping baseline in
$\ff(r;Q_{min})$, the two convolutions (thermal motion and termination effects)
are applied at different times during the PDF calculation.  The thermal
convolution is applied to $\rho_{bulk}(r)$ (\eq{rho}) and then this is converted
to $G(r)$ (\eq{grFromRhoBulk}), to which is applied the termination convolution.
Modeling programs~\cite{proff;jac99,farro;jpcm07} account for this correctly,
but when individual peaks are fit in real-space to extract peak positions and
intensities, this is often overlooked, though it seems reasonable to ignore the
termination effects in most cases.

\subsection{Calculating in real-space for nanoparticles modeled as attenuated bulk crystals}

In this case $\rho_{bulk} (r)$ is determined using a model of a bulk structure
as described in Section~\ref{sec:calcBulk}. The pair density,  $\rho (r)$, in
\eq{gofrgamma} is the function for the nanoparticle, which is approximated as
$\gamma_0 (r)\rho_{bulk} (r)$ \cite{guini;b;xdic63}.  Thus,
\begin{equation}
\label{eq:grnanofrombulk}
\begin{split}
\ff(r;Q_{min})&= 4\pi r \gamma_0 (r)(\rho_{bulk} (r) - \rho_0).
\end{split}
\end{equation}
This approach has been implemented in the PDFgui modeling software
\cite{farro;jpcm07} and used successfully on rather well ordered CdSe
nanocrystals \cite{masad;prb07}.  The main shortcoming is that effects that
cannot be incorporated in the average structure, such as surface relaxations or
core-shell inhomogeneities, cannot be modeled.

If there is a distribution of nanoparticle sizes and shapes, the characteristic
function, $\gamma_0(r)$ can be replaced with an appropriately averaged
characteristic function
\begin{equation}
\label{eq:gammaensemble}
\gamma(r)= \int  \gamma_0(r;R_1,R_2,\ldots) p(R_1,R_2,\ldots) \dd R_1\dd R_2
\ldots\,.
\end{equation}
Here,
$p(R_1,R_2,\ldots)$ is the normalized distribution of nanoparticle shapes parameterized by $R_1,\ R_2,\ldots$.  For example, for spherical nanoparticles of radius $R$, $p(R_1,R_2,\ldots)=p(R)$, the distribution of nanoparticle radii. Finally, we replace \eq{grnanofrombulk} with
\begin{equation}
\label{eq:PDFnano}
\begin{split}
\ff(r;Q_{min}) &= 4\pi r \gamma (r)(\rho_{bulk} (r) - \rho_0).
\end{split}
\end{equation}
Great care should be taken to ensure that the result is unique when refining a
number of nanoparticle morphology parameters beyond one or two.

\subsection{Calculating as the Fourier transform of the properly normalized Debye Function}

This approach has been successfully used by a number of authors
\cite{zhang;n03,cerve;jcc06}.  The $F(Q)$ function is evaluated using
\eq{fqdebye} and then Fourier transformed to obtain the desired real-space
function.  To account for thermal and zero-point motion in reciprocal-space
calculations, \eq{fqdebye} is replaced with a version that includes Debye-Waller
effects,
\begin{equation}
\label{eq:fqdebyewaller}
\begin{split}
F(Q) &= \frac{1}{N\langle f\rangle^2}\sum_{i\neq j} f_j^*f_i \left(e^{- \frac{1}{2}
\sigma_{ij}^2 Q^2}\right)\frac{ \sin (Qr_{ij})}{r_{ij}}.
\end{split}
\end{equation}
Here, $\sigma_{ij}^2$ is the correlated broadening factor for the atom pair,
which is the real-space Gaussian width discussed in the case of bulk
crystals.

As we show here, for a quantitative comparison with measured data, care must be
taken with the Fourier transform so that it is carried out over the same range
of $Q$ as the experiment.  The main drawback of the reciprocal-space approach is
that it can be very slow compared to direct real-space calculation due to the
long-range extent of the signal from each pair. It is generally preferred for
smaller systems.  However, due to recent algorithmic advances, the calculation
of the Debye equation for larger systems can be greatly accelerated under
certain circumstances~\cite{cerve;jcc06}.

Using this method, the termination effects coming from the Fourier transform are implicity included provided the Fourier transforms to obtain the model and data PDFs are terminated with the same $Q_{min}$ and $Q_{max}$ values.

\subsection{Calculating in real-space from discrete nanoparticle models}

In this case, \eq{gofrgamma} is used directly, where $\rho (r)$ is calculated
from a finite model of the discrete nanoparticle using \eq{rho}.  The difficulty
arises in determining a correct form for the baseline $-4\pi r\rho_0\gamma_0
(r)$.  Up until now, the shape of the baseline has been approximated using
expansions of \emph{ad hoc} mathematical
functions~\cite{korsu;jac05,neder;jpcm05,korsu;jac07,neder;pssc07}.  This is
successful at approximating the behavior of the baseline.  However, in this work
we derive the explicit form of the baseline shape in terms of the characteristic
function of the nanoparticle, the autocorrelation function of the nanoparticle
shape.  This suggests a number of approaches to calculate the PDF baseline in a
more physical way.

If we have accurate small angle scattering data from the samples, from
\eq{gofrgamma} we see that we can compute the PDF baseline from the measured SAS
via  a Fourier transform.  However, care must be exercised as the derivation
assumes that the sample is made up of discrete nanoparticles.  In general,
clusters and aggregates of nanoparticles will form and small angle scattering
signals from these structures on different length-scales will be present and
must be separated.  Also, scattering density fluctuations of any sort in the
sample will affect the SAS signal, as discussed elsewhere \cite{cargi;jac71}.
None of these effects need be explicitly considered if the SAS signal is not
used in the PDF definition, as in \eq{gofrgamma}, though an alternative method is
then required to determine the baseline.

The inclusion of SAS data has the potential to add
significant value to any refinement of nanoparticle models from the PDF. Both
small and large angle scattering contain information about the shape and size of
nanoparticles, but this information is nearly decoupled from the internal
nanoparticle structure in the small angle scattering. This same information is
in the PDF, but it can be obscured by structure features, such as when the PDF
prematurely attenuates due to a complex or amorphous surface structure.  In this
case the nanoparticle size obtained from SAS will be larger than the apparent NP
size obtained from the PDF that reflects the size of the coherent core
structure.  Even without the inclusion of SAS data into PDF refinements, we can
can learn much from SAS analysis techniques. PDF nanoparticle refinements
usually start with a simple model that includes atomic positions restricted by a
shape.  Therefore, PDF analysis can benefit from the various {\it ab initio}
methods~\cite{sverg;aca91,chaco;bpj98} for determining the shape of a scatterer
from the SAS.

Without the use of the small angle intensity for determining $\gamma_0 (r)$, we
can consider approaches to determine it self-consistently from the model since
it is the autocorrelation of the particle shape which is directly available from
the model itself by determining a ``shrink-wrapping" of the atomistic model. On
the contrary, when the internal structure is well known, but the size and shape
distribution of nanoparticles is not, then the characteristic function can be
parameterized and refined to obtain the approximate nanoparticle dimensions as
was done in \cite{masad;prb07}.

\section{The extent of small angle scattering}
\label{sec:sasft}

We have considered the two asymptotic situations here of including or excluding
all the SAS.  If the SAS is retained in the intensity that is Fourier
transformed, the resulting real-space function obtained is $4\pi
r\rho(r)=R(r)/r$ (\eq{frdebye}).  Excluding it all results in $G(r)$ (\eq{gofrgamma}).
We now consider the possibility that some, but not all, of the small angle scattering
is included in the Fourier transform.  This might occur in the
case of very small nanoparticles, for example, when the SAS extends to wider angles.
Here we estimate the circumstances under which a significant amount of small angle
intensity will appear in a wide angle PDF experiment for the case of a sphere of uniform density.
The scattering intensity is given by~\cite{rayle;prsl14}
\begin{equation}
\label{eq:firstpeak}
I(Q) \propto \frac{9}{(QR)^6} \left[\sin(QR) - QR\cos(QR)\right]^2,
\end{equation}
where $R$ is the radius of the sphere. By integrating this equation from 0 to
$Q_{min}$, and dividing by the total integrated intensity, we get an expression
for the proportion of small angle intensity above $Q_{min}$ for a given
nanoparticle diameter~\cite{rayle;prsl14}:
\begin{equation}
\begin{split}
\label{eq:integratedSphereSAS}
i(x)    &= 1- \frac{1}{2\pi r^5} \left[ \right.
             (2x^4-x^2+3)\cos(2x) \\
             &+ x(x^2+6)\sin(2x)
             + 4x^5\mathrm{Si}(2x) - (5x^2+3)
             \left. \right].
\end{split}
\end{equation}
Here, $x = Q_{min}R$ and $\mathrm{Si}(x)$ represents the sine integral,
$\mathrm{Si}(x) = \int_0^x\frac{\sin(x')}{x'}\dd x'$. For a very small
nanoparticle of radius $5 \mAA$, we see that $i(5) < 0.01$, corresponding to
$Q_{min} \approx 1 \mAA^{-1}$, which is typical for a RAPDF~\cite{chupa;jac03}
experiment.  Thus, for even quite small nanoparticles, practically all small
angle intensity is below $Q_{min}=1$\AA$^{-1}$ and \eq{gofrgamma} is
appropriate.  However, care should be taken not to extend $Q_{min}$ too low in
$Q$ in a measurement of a nanoparticulate system.

In the few-atom limit, such as the case of discrete small molecules,
the small and wide angle scattering are not cleanly
separated.  To produce a complete real-space signal, one can approximate the
small angle scattering from a candidate structure model. This approach is
commonly used in the study of small molecules in the gas
phase~\cite{hargi;b;saogped88}.  The Fourier transform of the estimated
scattering approximates $R(r)/r$, the nominal ``experimental'' or ``modified''
RDF. An equivalent method for obtaining the modified RDF from wide angle
scattering alone is to add a baseline estimated from a model structure to
$\ff(r, Q_{min})$ \cite{ruan;nanol07}.  This approach has been successful for
calculating the $R(r)/r$ for many-atom nanoparticles.

\section{Summary}
\label{sec:summary}

The PDF is a valuable tool for identifying the form and the interior composition
of nanoscale materials. Whereas the oscillating component of the PDF gives
information about the interatomic distances within the material, the PDF
baseline is a function of the characteristic function, a measure of nanoparticle shape
which has its origin in the SAS that is usually disregarded in a powder
diffraction experiment.  This characteristic function goes unnoticed in macroscopic
particles, where the PDF is observed at distances that are much smaller than the
particle diameter. For nanoparticles, the PDF baseline, and therefore the
characteristic function, cannot be disregarded.  We have presented a full
derivation of the PDF equation taking into account the missing SAS and have
reviewed different methods for calculating the PDF for nanoparticles. Given the
relationship between the PDF and SAS equations, there is potential benefit in
incorporating SAS data and analysis methods into PDF studies.

\section{Acknowledgments}

The authors would like to thank Phil Duxbury and Pavol Juh\'as for careful
proofreading and discussion of this manuscript.  The authors acknowledge
enlightening conversations with Matteo Leoni, Reinhard Neder, Thomas Proffen,
Chong-Yu Ruan, Paolo Scardi and Mike Thorpe. This work was supported by the US National
Science foundation through Grant DMR-0703940.

\bibliographystyle{nar}
\bibliography{abb-billinge-group,everyone,billinge-group}

\begin{thebibliography}{10}

\bibitem{debye;pz30}
Debye, P. and Menke, H. (1930)
{\em Physik. Z.} {\bf 31}, 797--8.

\bibitem{warre;b;xd90}
Warren, B.~E. (1990)
X-ray diffraction,
Dover, New York.

\bibitem{egami;b;utbp03}
Egami, T. and Billinge, S. J.~L. (2003)
Underneath the Bragg peaks: structural analysis of complex materials,
Pergamon Press, Elsevier, Oxford, England.

\bibitem{warre;jpc34}
Warren, B.~E. (1934)
{\em J. Phys. Chem.} {\bf 2}, 551.

\bibitem{warre;jacers36}
Warren, B.~E., Krutter, H., and Morningstar, O. (1936)
{\em J. Am. Ceram. Soc.} {\bf 19}, 202--6.

\bibitem{frank;ac50}
Franklin, R.~E. (1950)
{\em Acta Crystallogr.} {\bf 3}, 107.

\bibitem{frank;prsla51}
Franklin, R.~E. (1951)
{\em Proc. R. Soc. Lond.~A} {\bf 209}, 196.

\bibitem{wrigh;gpc98}
Wright, A.~C. (1998)
{\em Glass. Phys. Chem.} {\bf 24}, 148--179.

\bibitem{billi;cc04}
Billinge, S. J.~L. and Kanatzidis, M.~G. (2004)
{\em Chem. Commun.} {\bf 2004}, 749--760.

\bibitem{billi;s07}
Billinge, S. J.~L. and Levin, I. (2007)
{\em Science} {\bf 316}, 561--565.

\bibitem{mcken;n92}
McKenzie, D.~R., Davis, C.~A., Cockayne, D. J.~H., Muller, D.~A., and Vassallo,
  A.~M. FEB 13 1992
{\em Nature} {\bf 355(6361)}, 622--624.

\bibitem{zhang;n03}
Zhang, H.~Z., Gilbert, B., Huang, F., and Banfield, J.~F. (2003)
{\em Nature} {\bf 424(6952)}, 1025--1029.

\bibitem{gates;jpcb04}
Gateshki, M., Hwang, S.-J., Park, D., Ren, Y., and Petkov, V. (2004)
{\em J. Phys. Chem.~B} {\bf 108}, 14956--14963.

\bibitem{gilbe;s04}
Gilbert, B., Huang, F., Zhang, H., Waychunas, G.~A., and Banfield, J.~F. (2004)
{\em Science} {\bf 305}, 651--654.

\bibitem{page;chpl04}
Page, K., Proffen, T., Terrones, H., Terrones, M., Lee, L., Yang, Y., Stemmer,
  S., Seshadri, R., and Cheetham, A.~K. (2004)
{\em Chem. Phys. Lett.} {\bf 393}, 385--388.

\bibitem{korsu;jac05}
Korsunskiy, V.~I. and Neder, R.~B. Dec 2005
{\em J. Appl. Crystallogr.} {\bf 38(6)}, 1020--1027.

\bibitem{petko;jmc05}
Petkov, V., Gateshki, M., Choi, J., Gillan, E.~G., and Ren, Y. (2005)
{\em J. Mater. Chem.} {\bf 15}, 4654.

\bibitem{bedfo;jpcc07}
Bedford, N., Dablemont, C., Viau, G., Chupas, P., and Petkov, V. (2007)
{\em J. Phys. Chem.~C} {\bf 111(49)}, 18214--18219.

\bibitem{ehm;pd07}
Ehm, L., Antao, S.~M., Chen, J., Locke, D.~R., Michel, F.~M., Martin, C.~D.,
  Yu, T., Parise, J.~B., Antao, S.~M., Lee, P.~L., Chupas, P.~J., Shastri,
  S.~D., and Guo, Q. (2007)
{\em Powder Diffr.} {\bf 22}, 108--112.

\bibitem{masad;prb07}
Masadeh, A.~S., {Bo\v zin}, E.~S., Farrow, C.~L., Paglia, G., Juh\'as, P.,
  Karkamkar, A., Kanatzidis, M.~G., and Billinge, S. J.~L. (2007)
{\em Phys. Rev. B} {\bf 76}, 115413.

\bibitem{pradh;cm07}
Pradhan, S.~K., Mao, Y., Wong, S.~S., Chupas, P., and Petkov, V. (2007)
{\em Chem. Mater.} {\bf 19}, 6180--6186.

\bibitem{kaplo;pr65}
Kaplow, R., Strong, S.~L., and Averbach, B.~L. (1965)
{\em Phys. Rev.} {\bf 138}, 1336.

\bibitem{kaplo;pr68}
Kaplow, R., Rowe, T.~A., and Averbach, B.~L. (1968)
{\em Phys. Rev.} {\bf 168}, 1068.

\bibitem{klug;b;xdpfpaam74}
Klug, H.~P. and Alexander, L.~E. (1974)
X-ray diffraction procedures for polycrystalline and amorphous materials,
Wiley, New York 2nd edition.

\bibitem{johns;jncs82}
Johnson, P. A.~V., Wright, A.~C., and Sinclair, R.~N. (1982)
{\em J. Non-Cryst. Solids} {\bf 50}, 281--311.

\bibitem{soper;prl82}
Soper, A.~K. and Silver, R.~N. (1982)
{\em Phys. Rev. Lett.} {\bf 49(7)}, 471--474.

\bibitem{korsu;jsc85}
Korsunskii, V.~I. (1985)
{\em J. Struct. Chem.} {\bf 26}, 208--216.

\bibitem{wrigh;jncs85}
Wright, A.~C. (1985)
{\em J. Non-Cryst. Solids} {\bf 76}, 187.

\bibitem{nanao;prb87}
Nanao, S., Dmowski, W., Egami, T., Richardson, J.~W., and Jorgensen, J.~D.
  (1987)
{\em Phys. Rev.~B} {\bf 35(2)}, 435--440.

\bibitem{egami;ferro91}
Egami, T., Rosenfeld, H.~D., Toby, B.~H., and Bhalla, A. (1991)
{\em Ferroelectrics} {\bf 120}, 11.

\bibitem{billi;prl94}
Billinge, S. J.~L., Kwei, G.~H., and Takagi, H. (1994)
{\em Phys. Rev. Lett.} {\bf 72}, 2282.

\bibitem{petko;prb98}
Petkov, V., Gerber, T., and Himmel, B. (1998)
{\em Phys. Rev.~B} {\bf 58(18)}, 11982--11989.

\bibitem{petko;jac98}
Petkov, V. and Danev, R. (1998)
{\em J. Appl. Crystallogr.} {\bf 31}, 609.

\bibitem{keen;jac01}
Keen, D.~A. (2001)
{\em J. Appl. Crystallogr.} {\bf 34}, 172--177.

\bibitem{tucke;jac01a}
Tucker, M.~G., Dove, M.~T., and Keen, D.~A. (2001)
{\em J. Appl. Crystallogr.} {\bf 34}, 630--638.

\bibitem{soper;jpcm07ii}
Soper, A.~K. (2007)
{\em J. Phys.: Condens. Mat.} {\bf 19}, 335206.

\bibitem{neder;pssc07}
Neder, R.~B., Korsunskiy, V.~I., Chory, C., Müller, G., Hofmann, A., Dembski,
  S., Graf, C., and R\"uhl, E. (2007)
{\em Phys. Status Solidi~C} {\bf 4(9)}, 1610--1634.

\bibitem{billi;jssc08}
Billinge, S. J.~L. (2008)
{\em J. Solid State Chem.} {\bf 181}, 1698--1703.

\bibitem{dinne;b;pdtap08}
Robert E. Dinnebier and Simon J. L. Billinge, (ed.) (2008)
Powder diffraction: theory and practice,
Royal Society of Chemistry, London, England.

\bibitem{gilbe;jac08}
Gilbert, B. (2008)
{\em J. Appl. Crystallogr.} {\bf 41}, 554.

\bibitem{levas;jcc07}
Levashov, V.~A., Billinge, S. J.~L., and Thorpe, M.~F. (2007)
{\em J. Comput. Chem.} {\bf 28}, 1865--1882.

\bibitem{guini;b;xdic63}
Guinier, A. (1963)
X-ray diffraction in crystals, imperfect crystals, and amorphous bodies.,
San Francisco, W.H. Freeman, San Francisco.

\bibitem{rayle;prsl14}
Rayleigh, L. (1914)
{\em Proc. R. Soc. Lond} {\bf A90}, 219.

\bibitem{glatt;b;saxs82}
O. Glatter and O. Kratky, (ed.) (1982)
Small Angle X-ray Scattering,
Academic Press Inc., London 1st edition.

\bibitem{thorp;unpub07}
Thorpe, M.~F. and Lei, M.
Nanoparticle shape factors for spheroids
unpublished (2007).

\bibitem{kodam;aca06}
Kodama, K., Iikubo, S., Taguchi, T., and Shamoto, S. (2006)
{\em Acta Crystallogr.~A} {\bf 62}, 444--453.

\bibitem{qiu;prl05}
Qiu, X., {Th.~Proffen}, Mitchell, J.~F., and Billinge, S. J.~L. (2005)
{\em Phys. Rev. Lett.} {\bf 94}, 177203.

\bibitem{farro;jpcm07}
Farrow, C.~L., Juh\'as, P., Liu, J.~W., Bryndin, D., {Bo\v zin}, E.~S., Bloch,
  J., Proffen, T., and Billinge, S. J.~L. (2007)
{\em J. Phys: Condens. Mat.} {\bf 19}, 335219.

\bibitem{lizma;l96}
{Liz-Marz\'an}, L.~M., Giersig, M., and Mulvaney, P. (1996)
{\em Langmuir} {\bf 12}, 4329--4335.

\bibitem{neder;jpcm05}
Neder, R.~B. and Korsunskiy, V.~I. (2005)
{\em J. Phys.: Condens. Mat.} {\bf 17(5)}, S125--S134.

\bibitem{korsu;jac07}
Korsunskiy, V.~I., Neder, R.~B., Hofmann, A., Dembski, S., Graf, C., and
  R\"uhl, E. Dec 2007
{\em J. Appl. Crystallogr.} {\bf 40(6)}, 975--985.

\bibitem{debye;ap15}
Debye, P. (1915)
{\em Annalen der Physik (Berlin, Germany)} {\bf 46}, 809--823.

\bibitem{toby;aca92}
Toby, B.~H. and Egami, T. (1992)
{\em Acta Crystallogr.~A} {\bf 48(3)}, 336--46.

\bibitem{guini;b;sas55}
Guinier, A., Fournet, G., Walker, C., and Yudowitch, K. (1955)
Small-angle scattering of x-rays,
John Wiley \& Sons, Inc., New York.

\bibitem{billi;b;lsfd98}
Billinge, S. J.~L. (1998)
In S. J. L. Billinge and M. F. Thorpe, (ed.), Local Structure from Diffraction,
   New York: Plenum.
p. 137.

\bibitem{proff;jac99}
Proffen, T. and Billinge, S. J.~L. (1999)
{\em J. Appl. Crystallogr.} {\bf 32}, 572--575.

\bibitem{mcgre;ms88}
McGreevy, R.~L. and Pusztai, L. (1988)
{\em Mol. Simul.} {\bf 1}, 359--367.

\bibitem{thorp;b;fstpbas02}
Thorpe, M.~F., Levashov, V.~A., Lei, M., and Billinge, S. J.~L. (2002)
In S. J. L. Billinge and M. F. Thorpe, (ed.), From semiconductors to proteins:
  beyond the average structure,  New York: Kluwer/Plenum.
pp. 105--128.

\bibitem{jeong;jpc99}
{I.-K.~Jeong}, Proffen, T., Mohiuddin-Jacobs, F., and Billinge, S. J.~L. (1999)
{\em J. Phys. Chem.~A} {\bf 103}, 921--924.

\bibitem{jeong;prb03}
Jeong, I.~K., Heffner, R.~H., Graf, M.~J., and Billinge, S. J.~L. (2003)
{\em Phys. Rev. B} {\bf 67}, 104301.

\bibitem{cerve;jcc06}
Cervellino, A., Giannini, C., and Guagliardi, A. (2006)
{\em J. Comput. Chem.} {\bf 27}, 995--1008.

\bibitem{cargi;jac71}
Cargill, III, G.~S. (1971)
{\em J. Appl. Crystallogr.} {\bf 4}, 277.

\bibitem{sverg;aca91}
Svergun, D.~I. and Stuhrmann, H.~B. (1991)
{\em Acta Crystallogr.~A} {\bf 47(6)}, 736--744.

\bibitem{chaco;bpj98}
Chac\'on, P., Mor\'an, F., D\'iaz, J.~F., Pantos, E., and Andreu, J.~M. (1998)
{\em Biophys. J.} {\bf 74(6)}, 2760--2775.

\bibitem{chupa;jac03}
Chupas, P.~J., Qiu, X., Hanson, J.~C., Lee, P.~L., Grey, C.~P., and Billinge,
  S. J.~L. (2003)
{\em J. Appl. Crystallogr.} {\bf 36}, 1342--1347.

\bibitem{hargi;b;saogped88}
Istv\'an Hargittai and Magdolna Hargittai, (ed.) (1988)
Stereochemical applications of gas-phase electron diffraction,
VHC Publishers, New York 1st edition.

\bibitem{ruan;nanol07}
Ruan, C.-Y., Murooka, Y., Raman, R., and Murdick, R. (2007)
{\em Nano Lett.} {\bf 7(5)}, 1290--1296.

\end{thebibliography}
\end{document}